\documentclass[apj]{emulateapj}
\usepackage{psfig}
\usepackage{graphicx}
\usepackage{color}
\usepackage{subfloat}

\def\simlt{\lower.5ex\hbox{$\; \buildrel < \over \sim \;$}}
\def\simgt{\lower.5ex\hbox{$\; \buildrel > \over \sim \;$}}

\def\hnot{\ifmmode H_0 \else H$_0$ \fi}

\def\msun{\ifmmode M_{\odot} \else $M_\odot$\fi}
\def\lsun{\ifmmode L_{\odot} \else $L_\odot$\fi}
\def\kms{km s$^{-1}$}
\def\deg{\ifmmode ^{\circ}
 \else $^{\circ}$\fi}
\def\pdeg{\ifmmode
 $\setbox0=\hbox{$^{\circ}$}\rlap{\hskip.11\wd0 .}$^{\circ}
 \else \setbox0=\hbox{$^{\circ}$}\rlap{\hskip.11\wd0 .}$^{\circ}$\fi}
\def\msunyr{\ifmmode {\rm M_\odot~yr^{-1}}\else${\rm M_\odot~yr^{-1}}$\fi}
\def\lam{\ifmmode {\lambda} \else {$\lambda$} \fi}
\def\lamLlam{\ifmmode \lambda L_{\lambda}(5100) \else {$\lambda L_{\lambda}(5100)$} \fi}
\def\nuLnu{\ifmmode \nu L_{\nu}(5100) \else {$\nu L_{\nu}(5100)$} \fi}

\def\mdoto{\ifmmode {\dot{M}_0} \else $\dot{M}_0$ \fi}
\def\teff{\ifmmode {T_{eff}} \else $T_{eff}$ \fi}
\def\ilam{\ifmmode {I_\lambda} \else $I_\lambda$ \fi}
\def\flam{\ifmmode {F_\lambda} \else $F_\lambda$ \fi} 
\def\inu{\ifmmode {I_\nu} \else $I_\nu$ \fi}

\def\fnu{\ifmmode {F_\nu} \else $F_\nu$ \fi}
\def\yr{\ifmmode {\rm yr} \else yr \fi}
\def\cm{\ifmmode {\rm cm} \else cm \fi}
\def\cmmitwo{\ifmmode \rm cm^{-2} \else $\rm cm^{-2}$\fi}
\def\cmmithree{\ifmmode \rm cm^{-3} \else $\rm cm^{-3}$\fi}
\def\cmps{\ifmmode \rm cm~s^{-1}\else $\rm cm~s^{-1}$\fi}
\def\cmpsps{\ifmmode \rm cm~s^{-2}\else $\rm cm~s^{-2}$\fi}
\def\kmps{\ifmmode \rm km~s^{-1}\else $\rm km~s^{-1}$\fi}
\def\kmpspmpc{\ifmmode \rm km~s^{-1}~Mpc^{-1} \else
 $\rm km~s^{-1}~Mpc^{-1}$\fi}
\def\ergps{\ifmmode \rm erg~s^{-1} \else $\rm erg~s^{-1}$\fi}
\def\ergpspcm{\ifmmode \rm erg~s^{-1}~cm^{-2} \else $\rm erg~s^{-1}~cm^{-2}$ \fi}
\def\ergpspcmphz{\ifmmode \rm erg~s^{-1}~cm^{-2}~Hz^{-1} \else $\rm
erg~s^{-1}~cm^{-2}~Hz^{-1}$ \fi}
\def\ergpspcmpa{\ifmmode \rm erg~s^{-1}~cm^{-2}~\AA^{-1} \else $\rm
erg~s^{-1}~cm^{-2}~\AA^{-1}$ \fi}
\def\ergpsphz{\ifmmode \rm erg s^{-1} Hz^{-1} \else
 $\rm erg s^{-1} Hz^{-1}$ \fi}
\def\mbh{\ifmmode M_{\mathrm{BH}} \else $M_{\mathrm{BH}}$\fi}
\def\msigma{\ifmmode M_{\sigma} \else $M_{\sigma}$\fi}
\def\mbulge{\ifmmode M_{\mathrm{bulge}} \else $M_{\mathrm{bulge}}$\fi}
\def\lbulge{\ifmmode L_{\mathrm{bulge}} \else $L_{\mathrm{bulge}}$\fi}
\def\mgal{\ifmmode M_{\mathrm{gal}} \else $M_{\mathrm{gal}}$\fi}
\def\lsph{\ifmmode L_{\mathrm{sph}} \else $L_{\mathrm{sph}}$\fi}
\def\lgal{\ifmmode L_{\mathrm{gal}} \else $L_{\mathrm{gal}}$\fi}
\def\lgalbar{\ifmmode \bar{L}_{\mathrm{gal}} \else $\bar{L}_{\mathrm{gal}}$\fi}
\def\ltot{\ifmmode L_{\mathrm{tot}} \else $L_{\mathrm{tot}}$\fi}
\def\lfib{\ifmmode L_{\mathrm{fib}} \else $L_{\mathrm{fib}}$\fi}
\def\lfibbar{\ifmmode \bar{L}_{\mathrm{fib}} \else $\bar{L}_{\mathrm{fib}}$\fi}
\def\reff{\ifmmode {R_{e}} \else $R_{e}$ \fi}
\def\mgalstar{\ifmmode M^*_{\mathrm{gal}} \else $M^*_{\mathrm{gal}}$\fi}

\def\mbhsigstar{\ifmmode M_{\mathrm{BH}} - \sigma_* \else $M_{\mathrm{BH}} - \sigma_*$ \fi}
\def\mbhmbulge{\ifmmode M_{\mathrm{BH}} - M_{\mathrm{bulge}} \else $M_{\mathrm{BH}} - M_{\mathrm{bulge}}$ \fi}
\def\mbhlgal{\ifmmode M_{\mathrm{BH}} - M_{\mathrm{gal}} \else $M_{\mathrm{BH}} - L_{\mathrm{gal}}$ \fi}
\def\mbhmgal{\ifmmode M_{\mathrm{BH}} - L_{\mathrm{gal}} \else $M_{\mathrm{BH}} - M_{\mathrm{gal}}$ \fi}
\def\dmbh{\ifmmode \Delta~{\mathrm{log}}~M_{\mathrm{BH}} \else $\Delta$~log~$M_{\mathrm{BH}}$\fi}

\def\sigstar{\ifmmode \sigma_* \else $\sigma_*$\fi}
\def\signl{\ifmmode \sigma_{\mathrm{NL}} \else $\sigma_{\mathrm{NL}}$\fi}
\def\sigthree{\ifmmode \sigma_{\mathrm{[O~III]}} \else $\sigma_{\mathrm{[O~III]}}$\fi}
\def\sigtwo{\ifmmode \sigma_{\mathrm{[O~II]}} \else $\sigma_{\mathrm{[O~II]}}$\fi}
\def\wthree{\ifmmode {\rm FWHM({[O~III]})} \else $FWHM({[O~III]})$ \fi}
\def\wtwo{\ifmmode {\rm FWHM({[O~II]})} \else $FWHM({[O~II]})$ \fi}
\def\mthree{\ifmmode M_{\mathrm [O~III]} \else $M_{\mathrm [O~III]}$ \fi}
\def\mtwo{\ifmmode M_{\mathrm [O II]} \else $M_{\mathrm [O II]}$ \fi}
\def\hbeta{\ifmmode {\rm H}\beta \else H$\beta$\fi}
\def\hdelta{\ifmmode {\rm H}\delta \else H$\delta$\fi}
\def\hepsilon{\ifmmode {\rm H}\epsilon \else H$\epsilon$\fi}
\def\mgii{\ifmmode {\rm Mg{\sc ii}} \else Mg~{\sc ii}\fi}

\def\lbreak{\ifmmode L_{\mathrm{break}} \else $L_{\mathrm{break}}$\fi}
\def\lcut{\ifmmode L_{\mathrm{cut}} \else $L_{\mathrm{cut}}$\fi}
\def\led{\ifmmode L_{\mathrm{Edd}} \else $L_{\mathrm{Edd}}$\fi}
\def\lbol{\ifmmode L_{\mathrm{bol}} \else $L_{\mathrm{bol}}$\fi}

\newcommand{\oiii}{{\sc [O~iii]}}

\newcommand{\caii}{Ca~{\sc ii}~H$+$K}
\newcommand{\neiii}{[Ne~{\sc iii}]}

\shorttitle{Quasar Hosts}
\shortauthors{Salviander et al.}

\begin{document}

\title{THE BLACK HOLE MASS--GALAXY LUMINOSITY RELATIONSHIP FOR SLOAN DIGITAL SKY SURVEY QUASARS}

\author{S. Salviander\altaffilmark{1}, G.~A. Shields\altaffilmark{1}, {\sc and} E.~W. Bonning\altaffilmark{2}}

\altaffiltext{1}{Department of Astronomy, University of Texas, Austin, TX 78712, USA; triples@astro.as.utexas.edu, shields@astro.as.utexas.edu}

\altaffiltext{2}{Department of Physics, Emory University, Atlanta, GA 30322, USA; erin.bonning@emory.edu}


\begin{abstract}

We investigate the relationship between the mass of the central supermassive black hole, \mbh, and the host galaxy luminosity, \lgal, in a sample of quasars from the Sloan Digital Sky Survey Data Release 7. We use composite quasar spectra binned by black hole mass and redshift to assess galaxy features that would otherwise be overwhelmed by noise in individual spectra. The black hole mass is calculated using the photoionization method, and the host galaxy luminosity is inferred from the depth of the \caii\ features in the composite spectra. We evaluate the evolution in the \mbh\ -- \lgal\ relationship by examining the redshift dependence of \dmbh, the offset in \mbh\ from the local \mbh\ -- \lgal\ relationship. There is little systematic trend in \dmbh\ out to $z = 0.8$. Using the width of the \oiii\ emission line as a proxy for the stellar velocity dispersion, \sigstar, we find agreement of our derived host luminosities with the locally observed Faber--Jackson relation. This supports the utility of the width of the \oiii\ line as a proxy for \sigstar\ in statistical studies.
 
\end{abstract}


\section{INTRODUCTION}\label{s:intro}

The co-evolution of galaxies and their central black holes is a subject of intensive study. The relationship between the mass of the black hole, \mbh, and the properties of the host galaxy may hold clues to the physics of baryon assembly in galactic evolution and the back-reaction of active galactic nuclei (AGNs) on their host galaxies. For a recent review of the properties of galaxies and their black holes, see \citet{kormendy13}. It is clear that \mbh\ increases in rough proportion to the luminosity \lgal\ and mass \mgal\ of the bulge component of the host galaxy \citep[e.g.,][]{magorrian98} and to $\sigstar^4$\citep{ferrarese00, gebhardt00}. However, there are a number of outstanding issues regarding the linearity and scatter of the relationship over the full range of \mbh, and the nature of the relationship in the case of pseudo-bulges \citep{kormendy13}. Of great interest is the question of the evolution of the black hole--bulge relationship over cosmic time. Results to date tend to suggest smaller \lgal\ for a given \mbh\ at large redshift, but various studies have reached seemingly contradictory conclusions. New measurements using independent techniques are therefore of value.

Using a large sample of quasars from the Sloan Digital Sky Survey (SDSS)\footnote{The SDSS website is http://www.sdss.org.}, \citet[][hereinafter ``S13'']{salviander13} assessed the evolution of the \mbhsigstar\ relationship and found little change back to $z \approx 1.0$. That work used the width of the \oiii~$\lambda5007$ line as a surrogate for \sigstar. In this paper, we report results of a complementary study of the evolution of the \mbhlgal\ relationship for the same quasar sample. Here we use composite spectra to achieve a sufficient signal-to-noise ratio (S/N) to permit the measurement of the \caii\ absorption lines in the host galaxy starlight contained in the quasar spectra. This allows us to assess the evolution of the \mbhlgal\ relationship without recourse to the \oiii\ surrogacy.

For the sake of economy, we assume familiarity with S13, which gives background and references. We use cosmological parameters $\hnot = 70~\kmpspmpc, \Omega_{\rm M} = 0.3$, and $\Omega_{\Lambda} = 0.7$.

\section{SAMPLE AND METHOD}\label{s:method}

\subsection{Sample Selection and Spectrum Measurements}\label{s:selection}

Sample selection and spectral measurements are described in S13 and \citet[][hereinafter ``S07'']{salviander07}. The quasars in our sample were drawn from the SDSS Data Release 7 (Abazajian et al. 2009). We selected all spectra classified as quasars in the Spectroscopic Query Form in the redshift range $0.1 \leq z \leq 0.81$ in order to include both the \hbeta\ and \oiii\ emission lines at the highest possible redshift (the ``HO3'' sample described in S13). After imposing a series of quality cuts to remove substandard spectra, the final sample consists of 5355 individual quasars. 

\begin{figure*}
\epsscale{0.8}
\plotone{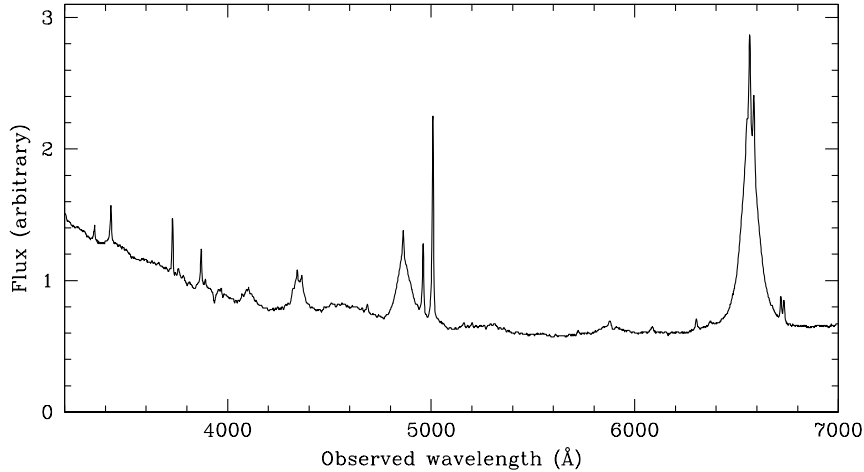}
\caption{Composite quasar spectrum for the ``8.25'' mass bin. See Table \ref{t:mbins} for data for this bin.}
\label{f:comp}
\end{figure*}

We calculated black hole mass for each of the quasars using the ``photoionization method,'' described in Section 2.1 of S13. The black hole mass is given by the equation
\begin{equation}\label{e:mbh}
\mbh = (10^{7.69} \ \msun) v_{3000}^2 L_{44}^{0.5},
\end{equation}
where $v_{3000}$ is the \hbeta\ broad line FWHM in units of 3000 \kms\ and $L_{44}$ is the 5100 \AA\ quasar luminosity in units of $10^{44}$ \ergps\ \citep{shields03}. This formula is adopted here for the sake of continuity with our earlier work (S07, S13). It is reasonably consistent with calibrations such as \citet{onken04}. We caution, however, that the analysis by \citet{kormendy13} gives a somewhat different slope and a substantially larger intercept in the \mbhlgal\ relationship for ellipticals and classical bulges, and finds that pseudo-bulges typically have smaller black holes and more scatter. These results provide a potential basis for a recalibration of expressions like Equation (\ref{e:mbh}).

Where needed (see Section \ref{s:faber}), we use the width of the quasar narrow \oiii\ emission line as a proxy for \sigstar, with $\signl =$ FWHM(\oiii)$/2.35$ for a Gaussian profile. There is considerable scatter in the correlation between \sigstar\ and \signl, but in the mean the two quantities track each other for a wide range of AGN luminosities \citep[e.g.,][]{bonning05,greene06oiii,shields06leiden}.

\subsection{Composite Spectra}\label{s:comp}

For the quasars studied here, the galaxy contributes $\sim 10\%$ to $30\%$ of the total observed continuum. Therefore, the stellar absorption features are weak and easily masked by the complex emission-line spectrum of the quasar. In these circumstances, the only identifiable stellar feature may be the \caii\ lines at rest wavelengths $\lambda3968$ and $\lambda3933$
\citep{greene06sig}. Even this feature typically is lost in the noise for the individual SDSS quasar spectra. However, the K line is clearly visible, for example, in the high S/N composite quasar spectrum of \citet{vandenberk01}. Therefore, we composed a set of composite spectra using subsets of our SDSS quasar sample, designed to permit study of the \mbhlgal\ relationship through the use of the H \& K lines.

The quasars in our sample were assigned to bins on the basis of black hole mass, \mbh, as described in S13. The mass bins range from $\mbh < 10^{7.0} \ \msun$ to $\mbh > 10^{9.0} \ \msun$, and are incremented by 0.5 dex \msun. We used an automated algorithm to create composite spectra from the individual SDSS spectra for the quasars in each bin. This program shifted the individual spectra to a common rest-wavelength scale running from 3200 to 8000~\AA\ with a linear spacing of 1.4~\AA. We used the redshift of the narrow \oiii\ lines for individual quasars derived from an automated spectrum-fitting algorithm \citep{salviander07}. The specific flux for each object was scaled to give a mean scaled flux density $f_\lambda$ of unity for the observed wavelength points in the rest wavelength interval $\lambda5100\pm20$~\AA; that is, $f_\lambda = \flam/x_c$, where $x_c$ is the mean of the observed \flam\ in this wavelength range. The composite flux density at a particular rest wavelength was then computed as the mean of the flux for all quasars contributing at that wavelength.

Figure \ref{f:comp} shows a composite quasar spectrum for the ``8.25'' mass bin (see Table \ref{t:mzbins} for data for this bin), which has a mean log \mbh $= 8.23$ \msun\ and mean $z = 0.436$. 

\begin{subfigures}
\begin{figure*}
\epsscale{0.8}
\plotone{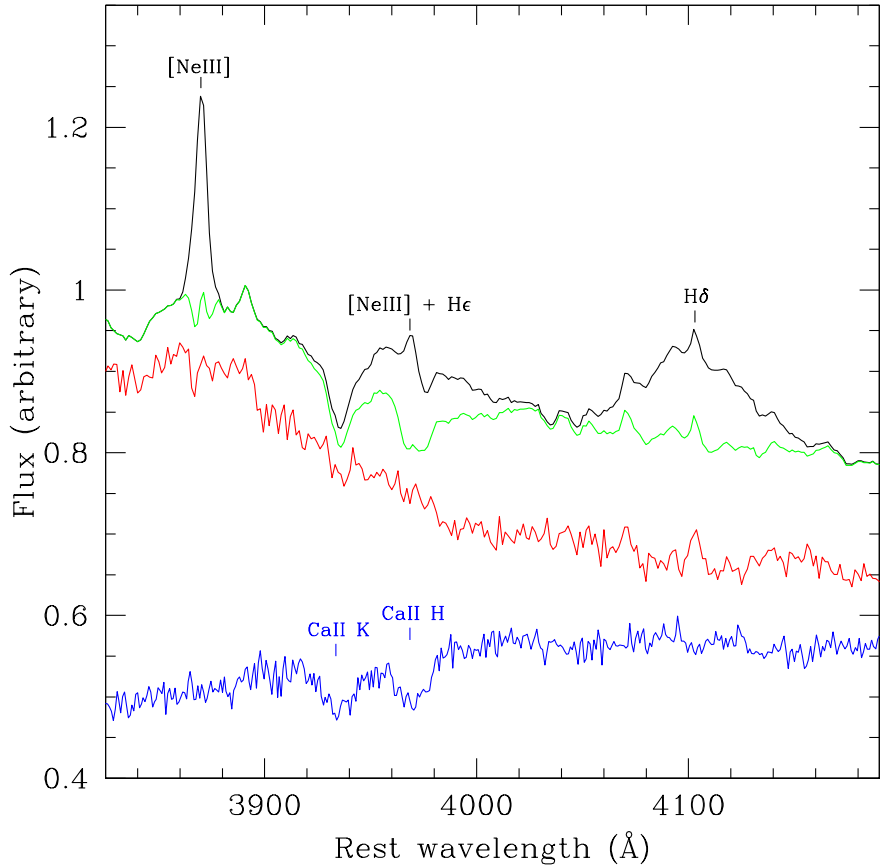}
\caption{Composite quasar spectrum for the ``8.25'' mass bin in the \caii\ region. The black line shows the spectrum; the green line shows the spectrum after subtraction of the \hepsilon\ and \neiii~$\lambda3967$ features; the red line shows the spectrum after subtraction of the scaled galaxy template (blue line).}\label{f:compsuba} 
\end{figure*}


\begin{deluxetable*}{ccccccccccccc}
\tabletypesize{\scriptsize}
\tablewidth{0pt}
\tablecaption{Average Quantities for the Mass Bins\label{t:mbins}}
\tablehead{
\colhead{Bin} &
\colhead{log \mbh} & 
\colhead{Size} &
\colhead{$z$} & 
\colhead{$d_L$} & 
\colhead{$x_c$} & 
\colhead{Factor} &
\colhead{$\lambda F_{\mathrm{\lambda, gal}}$} & 
\colhead{log $L_{\mathrm{gal}}$} & 
\colhead{log $L_{\mathrm{gal}}$} & 
\colhead{log $M_{\mathrm{gal}}$} & 
\colhead{log \signl} &
\colhead{\dmbh} \\
\colhead{} &
\colhead{(\msun)} &
\colhead{} &
\colhead{} &
\colhead{(Mpc)} &
\colhead{} &
\colhead{} &
\colhead{} &
\colhead{(\lsun, $V$)} &
\colhead{(pass. evol.)} &
\colhead{(pass. evol.)} &
\colhead{(\kmps)} &
\colhead{(pass. evol.)}}
\startdata
6.75	&	6.91	&	 71	&	0.170	&	 723.4	&	 6.10	&	 32	&	4.29	&	 9.95	&	 9.88	&	10.552	&	2.10	&	$-$0.79	 \\
7.25	&	7.29	&	 885		&	0.273	&	1154.5	&	 6.89	& 	 53	&	2.56	&	10.13	&	10.03	&	10.724	&	2.17	&	$-$0.60	 \\
7.75	&	7.75	&	1965		&	0.360	&	1520.5	&	 7.79	&	 58	&	2.35	&	10.36	&	10.24	&	10.968	& 	2.21	&	$-$0.42	 \\
8.25	&	8.21	&	1808		&	0.436	&	1958.9	&	 8.70	&	 71	&	1.93	&	10.55	&	10.39	&	11.154	&	2.24	&	$-$0.17	 \\
8.75	&	8.64	&	 520		&	0.509	&	2475.7	&	12.68	&	130	&	1.05	&	10.64	&	10.47	&	11.240	&	2.29	&	$+$0.18	 \\
9.25	&	9.12	&	 38		&	0.556	&	2793.4	&	24.12	&	347	&	3.94	&	10.63	&	10.44	&	11.213	&	2.31	&	$+$0.68	 \\
\enddata
\tablecomments{Log \mbh, $x_c$, $d_L$ are harmonic means; $z$ is an arithmetic mean. Units for $\lambda F_{\mathrm{\lambda, gal}}$ are $10^{-14} \ergpspcm$. See text for an explanation of the other quantities.}
\end{deluxetable*}

\begin{figure*}
\epsscale{0.7}
\plotone{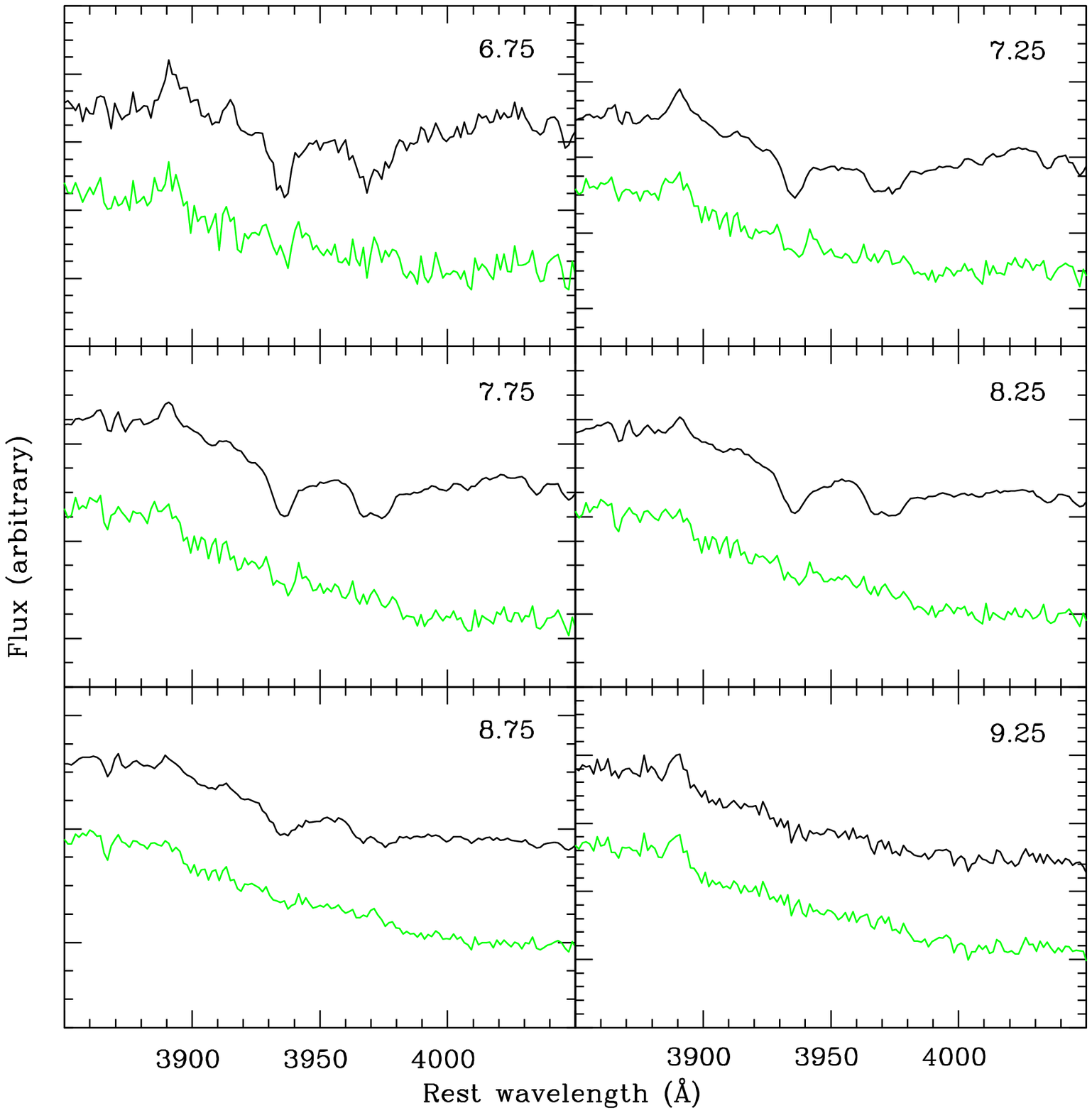}
\caption{Composite quasar spectra for all of the mass bins in the \caii\ region. The black line shows the spectrum after subtraction of the \hepsilon\ and \neiii~$\lambda3967$ features; the green line shows the spectrum after subtraction of the scaled galaxy template. See Table \ref{t:mbins} for data for these bins.}\label{f:compsubb} 

\epsscale{0.7}
\plotone{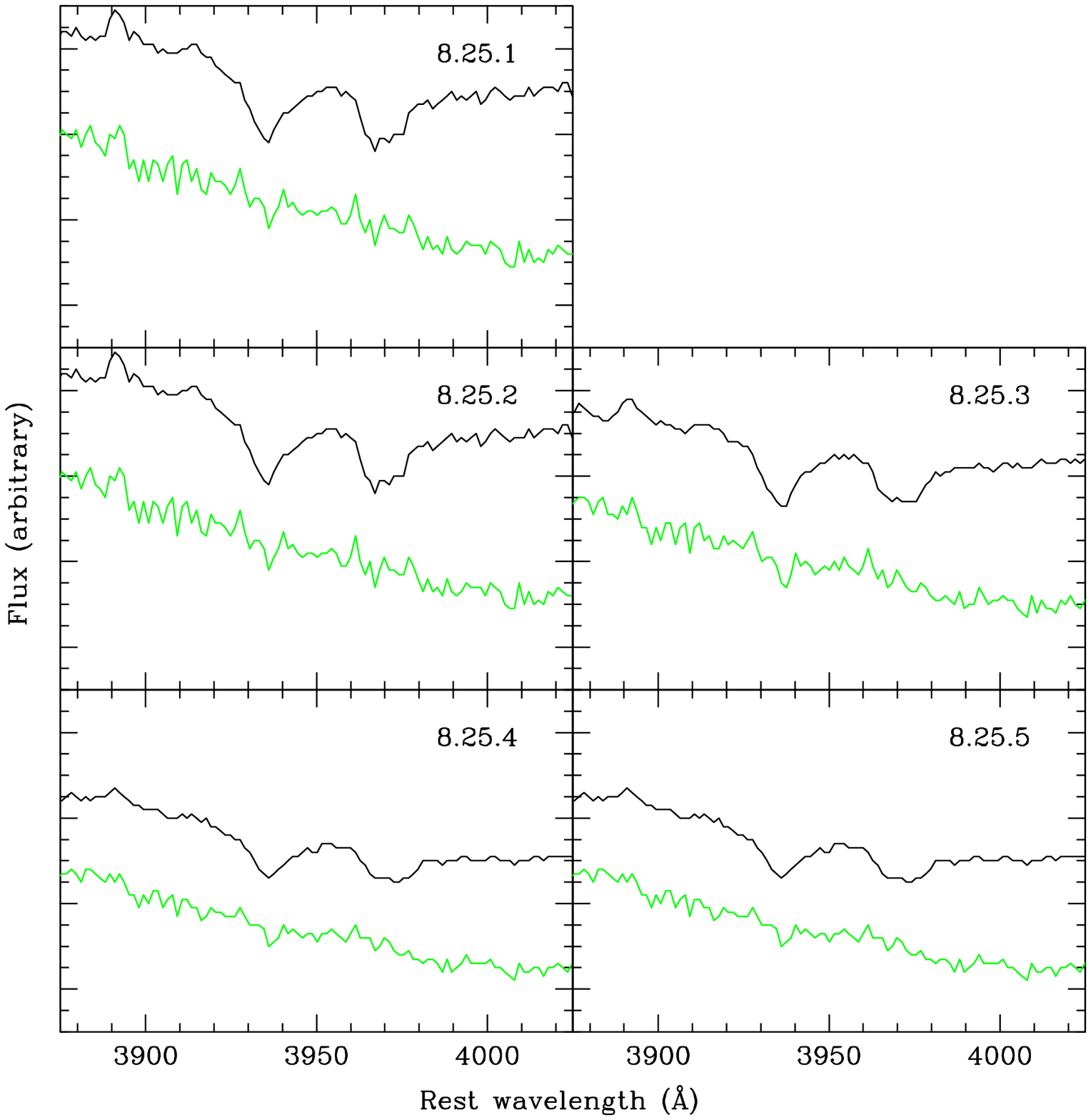}
\caption{Composite quasar spectrum for all of the ``8.25'' mass-redshift bins in the \caii\ region. The black line shows the spectrum after subtraction of the \hepsilon\ and \neiii~$\lambda3967$ features; the green line shows the spectrum after subtraction of the scaled galaxy template. See text and Table \ref{t:mzbins} for discussion and redshift values. }\label{f:compsubc}
\end{figure*}

\end{subfigures}

\subsection{Deriving Galaxy Luminosity}\label{s:lgal}

We derived the host galaxy luminosity for the quasars contributing to a given composite spectrum from the strength of the \caii\ absorption feature. This spectral region is heavily affected by the quasar emission-line spectrum. In particular, the narrow line of \neiii\ at $\lambda3967$ and the broad line of \hepsilon\ at $\lambda3970$ overlap the Ca H line. Following \citet{shields06co}, we subtracted the \neiii\ line using the theoretical ratio of $I(\lambda3967)/I(\lambda3869) = 0.31$ \citep{osterbr06} and the observed properties of the stronger $\lambda3869$ line. We also subtracted the broad \hepsilon\ line using a theoretical ratio $I(\hepsilon)/I(\hdelta) = 0.6$ \citep{shields06co} and the observed flux and profile of \hdelta. Figure \ref{f:compsuba} shows the result for the ``8.25'' composite spectrum, which now shows the \caii\ H line along with the K line.

In order to derive the host galaxy contribution to a given composite spectrum, we employed a template galaxy spectrum. We chose SDSS J151741.75-004217.6 (spectroscopic designation spSpec-51689-0312-142) at redshift $z = 0.1161$, an early type galaxy from the study of \citet{bernardi08}. This object has a good quality SDSS spectrum and a luminosity $L_V = 10^{10.30}~\lsun$ typical of the host galaxy luminosities found in the present work. This galaxy spectrum was scaled in flux and subtracted from the composite quasar spectrum in a trial-and-error fashion until the \caii\ lines were absent from the resulting spectrum on the basis of (1) visual inspection or (2) a least squares fitting procedure. The scaled flux density of the template spectrum gives the strength of the host galaxy contribution to the composite spectrum. Figure \ref{f:compsuba} illustrates a representative composite spectrum before and after subtraction of the scaled galaxy template spectrum.

The least-squares procedure involved a target continuum consisting of a straight line in \flam\ versus $\lambda$ anchored at two points defined by the average of \flam\ for the measured points in the intervals from 3910 to 3920~\AA\ and from 3980 to 3990~\AA. The scale factor for the galaxy spectrum was varied to minimize the mean square deviation of the galaxy-subtracted spectrum from the linear continuum in the wavelength range from 3920 to 3980~\AA. The reduced chi-squared for the numerical best fit was $\sim1.0$. Because of the complex nature of the quasar spectrum in this wavelength range, uncertainties were estimated by visually exploring the limiting values of the scale factor that failed to remove fully the H \& K lines (under-subtraction) or that produced a spurious emission feature at the wavelength of the H \& K lines (over-subtraction). The results quoted here are the galaxy flux from the least-squares fit together with error limits from the visual procedure, which are more conservative than the formal uncertainties in the least squares procedure. The typical difference between the least-squares results and the mid-point of the error limits serves as one indication of the uncertainties in the measurements.

Recovery of the characteristic host galaxy luminosity from the flux in a given composite spectrum requires the computation of suitable averages for the luminosity distance $d_L$ and the factor $x_c$ by which the individual spectra were divided to produce the composite. The considerations given in the Appendix lead to the expression
\begin{equation}\label{e:lbar}
\bar{L}_{\lambda,{\rm gal}} = f_{\lambda,{\rm gal}}^c \langle x_c \, (1+z)\,4\pi d_L^2\rangle_h.
\end{equation}
Here, $f_{\lambda,{\rm gal}}^c$ is the galaxy component of the flux density at wavelength $\lambda$ in the composite spectrum. Thus, we reverse the division by $x_c$ so as to return to a true flux scale, and we multiply by $4\pi d_L^2$. The factor $(1+z)$ is related to the definition of $d_L$. The subscript ``h'' denotes the harmonic mean of the quantity in angle brackets over the quasars contributing to the given composite spectrum. As discussed in the Appendix, the resulting luminosity is approximately the harmonic mean of the luminosities of the host galaxies for the quasars in the given composite spectrum, $\bar{L}_{\rm gal} = \langle 1/L_{\rm gal}\rangle^{-1}$. We converted our host luminosities to mass \mgal\ using Equation (10) of \citet{magorrian98} for $M/L$.

Table \ref{t:mbins} shows various measured and averaged quantities for the composite spectra for each mass bin. Figure \ref{f:compsubb} shows the subtracted spectra for the mass bins. Note the decreasing prominence of the calcium lines with increasing \mbh, reflecting a decreasing ratio of galaxy to AGN luminosity. For the 9.25 mass bin, the H \& K lines are barely visible, and the quoted value of \lgal\ might reasonably be treated as an upper limit.

\begin{figure}
\epsscale{1.2}
\plotone{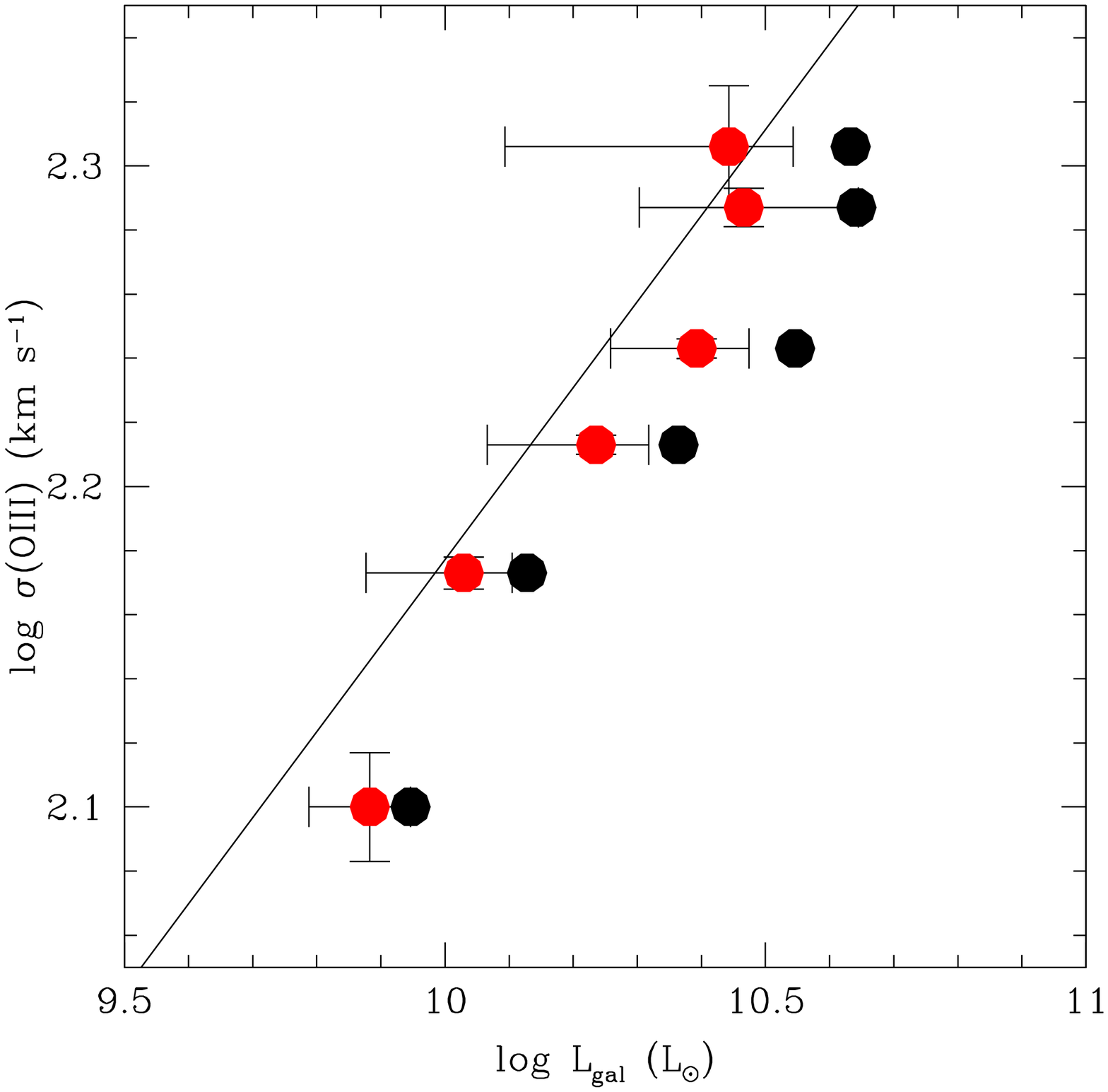}
\caption{Host galaxy luminosity vs. \signl\ for our sample. The red dots show the data after correction for passive evolution (see Section 3.1). The solid line is the Faber--Jackson relationship for local coreless elliptical galaxies \citep{kormendy13fj}.}\label{f:fj}
\end{figure}

\begin{figure}
\epsscale{1.2}
\plotone{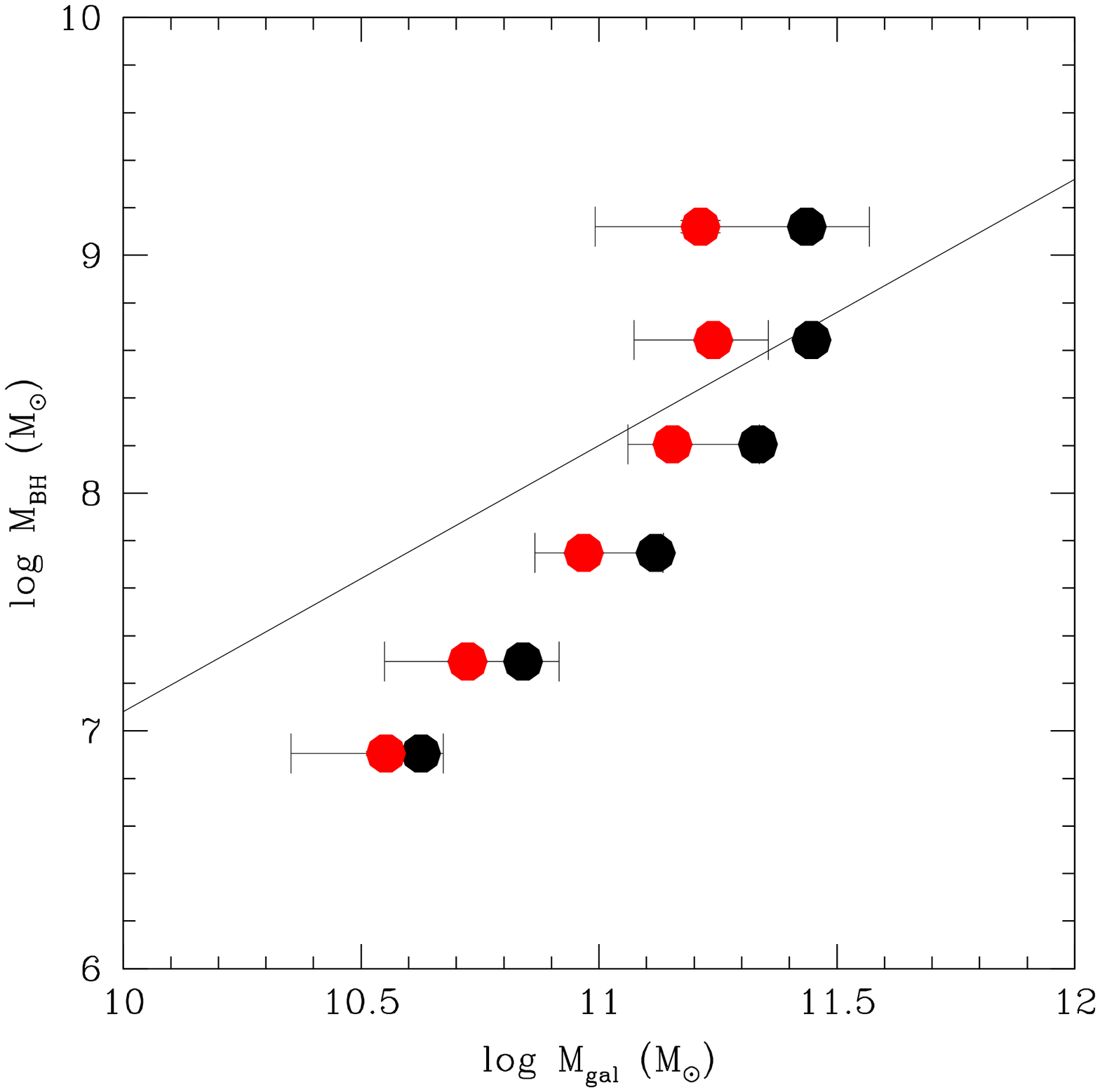}
\caption{\mbhmgal\ relationship for our sample. The red dots show the data after correction for passive evolution (see Section 3.1). The solid line is the \mbhmgal\ relationship for locally observed galaxies given by \citet{haering04}. See text and Table \ref{t:mbins} for the boundaries of the mass bins and other quantities; the size of the points exceeds the standard error of the mean for \mbh\ for the bins.} \label{f:mbhmgal}
\end{figure}

\section{RESULTS}\label{s:results}

\subsection{Faber--Jackson Relation}\label{s:faber}

Figure \ref{f:fj} shows the correlation between \signl\ and \lgal\ (visual) for the mass bins. Corrections for evolution of the stellar population were made with the aid of the Passive Evolution Calculator\footnote{http://www.astro.yale.edu/dokkum/evocalc/} of \citet{vandokkum01}, using Default 1 (stars formed at $z >> 1$). (Because $\bar{L}_{\rm gal}$ is a harmonic mean, we define a mean $\bar{\sigma}_\mathrm{NL} \equiv [\langle \signl^{-4}\rangle]^{-0.25}$; this would give agreement if all galaxies perfectly obeyed $\lgal \propto \sigstar^4$.) The solid line is the Faber--Jackson relation determined for nearby coreless elliptical galaxies \citep{kormendy13fj}. Our sample of quasar host galaxies generally follows the Faber--Jackson relation for the luminosity range involved, namely $9.9 \ \lsun < $ log \lgal\ $ < 10.7 \ \lsun$. The displacement toward larger host luminosity could involve a disk contribution to the host luminosity, especially for the lower luminosity bins (see Section \ref{s:disk}). The overall similarity of the trend in Figure \ref{f:fj} to the expected F-J slope supports the validity of the present technique for deriving the host galaxy luminosity. It also gives support to the use of \sigthree\ as a proxy for \sigstar\ in statistical samples of quasars.


\begin{deluxetable*}{ccccccccccccc}
\tabletypesize{\scriptsize}
\tablewidth{0pt}
\tablecaption{Average Quantities for the Mass Bins\label{t:mzbins}}
\tablehead{
\colhead{Bin} &
\colhead{log \mbh} & 
\colhead{Size} &
\colhead{$z$} & 
\colhead{$d_L$} & 
\colhead{$x_c$} & 
\colhead{Factor} &
\colhead{$\lambda F_{\mathrm{\lambda, gal}}$} & 
\colhead{log $L_{\mathrm{gal}}$} & 
\colhead{log $L_{\mathrm{gal}}$} & 
\colhead{log $M_{\mathrm{gal}}$} & 
\colhead{log \signl} &
\colhead{\dmbh} \\
\colhead{} &
\colhead{(\msun)} &
\colhead{} &
\colhead{} &
\colhead{(Mpc)} &
\colhead{} &
\colhead{} &
\colhead{} &
\colhead{(\lsun, $V$)} &
\colhead{(pass. evol.)} &
\colhead{(pass. evol.)} &
\colhead{(\kmps)} &
\colhead{(pass. evol.)}}
\startdata
6.75.1	&	6.89	&	 34	&	0.124	&	 570.3	&	11.18	&	 29.2	&	4.67	&	 9.91	&	 9.86	&	10.528	&	2.09	&	$-$0.78	 \\
6.75.2	&	6.92	&	 33	&	0.191	&	 903.3	&	 5.82	&	 33.5	&	4.07	&	 9.99	&	 9.92	&	10.599	&	2.11	&	$-$0.83	 \\
7.25.1	&	7.25	&	114	&	0.129	&	 597.5	&	15.15	&	 32.0	&	4.26	&	10.04	&	 9.99	&	10.684	&	2.14	&	$-$0.59	 \\
7.25.2	&	7.27	&	458	&	0.227	&	1084.3	&	 7.55	&	 44.0	&	3.10	&	10.15	&	10.07	&	10.773	&	2.17	&	$-$0.68	 \\
7.25.3	&	7.34	&	256	&	0.366	&	1918.4	&	 5.37	&	105.8	&	1.29	&	10.16	&	10.03	&	10.731	&	2.20	&	$-$0.56	 \\
7.75.1	&	7.74	&	 51	&	0.127	&	 583.1	&	18.08	&	 28.6	&	4.78	&	10.15	&	10.10	&	10.810	&	2.14	&	$-$0.25	 \\
7.75.2	&	7.72	&	118	&	0.232	&	1109.2	&	 9.49	&	 42.3	&	3.22	&	10.30	&	10.22	&	10.946	&	2.21	&	$-$0.42	 \\
7.75.3	&	7.75	&	630	&	0.376	&	1979.2	&	 7.10	&	 77.4	&	1.76	&	10.45	&	10.31	&	11.059	&	2.24	&	$-$0.52	 \\
7.75.4	&	7.78	&	706	&	0.514	&	2897.3	&	 6.35	&	166.0	&	0.82	&	10.44	&	10.26	&	11.000	&	2.21	&	$-$0.42	 \\
8.25.1	&	8.17	&	 43	&	0.128	&	 593.8	&	26.78	&	 38.2	&	3.57	&	10.19	&	10.14	&	10.859	&	2.18	&	$+$0.12	 \\
8.25.2	&	8.18	&	346	&	0.237	&	1144.2	&	12.00	&	 45.9	&	2.97	&	10.40	&	10.31	&	11.057	&	2.25	&	$-$0.09	 \\
8.25.3	&	8.20	&	624	&	0.376	&	1978.8	&	 7.85	&	 67.8	&	2.01	&	10.55	&	10.42	&	11.180	&	2.25	&	$-$0.20	 \\
8.25.4	&	8.21	&	531	&	0.516	&	2909.1	&	 7.94	&	130.0	&	1.05	&	10.64	&	10.46	&	11.233	&	2.25	&	$-$0.25	 \\
8.25.5	&	8.25	&	281	&	0.657	&	3906.9	&	 8.40	&	262.3	&	0.52	&	10.66	&	10.43	&	11.203	&	2.28	&	$-$0.18	 \\
8.75.2	&	8.63	&	 49	&	0.246	&	1178.9	&	16.79	&	 64.8	&	2.10	&	10.46	&	10.37	&	11.128	&	2.36	&	$+$0.29	 \\
8.75.3	&	8.64	&	142	&	0.383	&	2029.5	&	12.46	&	102.2	&	1.34	&	10.58	&	10.45	&	11.217	&	2.30	&	$+$0.19	 \\
8.75.4	&	8.64	&	163	&	0.526	&	2974.8	&	11.75	&	182.5	&	0.75	&	10.67	&	10.49	&	11.271	&	2.28	&	$+$0.13	 \\
8.75.5	&	8.66	&	147	&	0.673	&	4022.7	&	12.65	&	268.3	&	0.51	&	10.85	&	10.63	&	11.432	&	2.28	&	$-$0.03	 \\
\enddata
\tablecomments{Log \mbh, $x_c$, $d_L$ are harmonic means; $z$ is an arithmetic mean. Units for $\lambda F_{\mathrm{\lambda, gal}}$ are $10^{-14} \ergpspcm$. See text for an explanation of the other quantities.}
\end{deluxetable*}

\begin{figure}
\epsscale{1.2}
\plotone{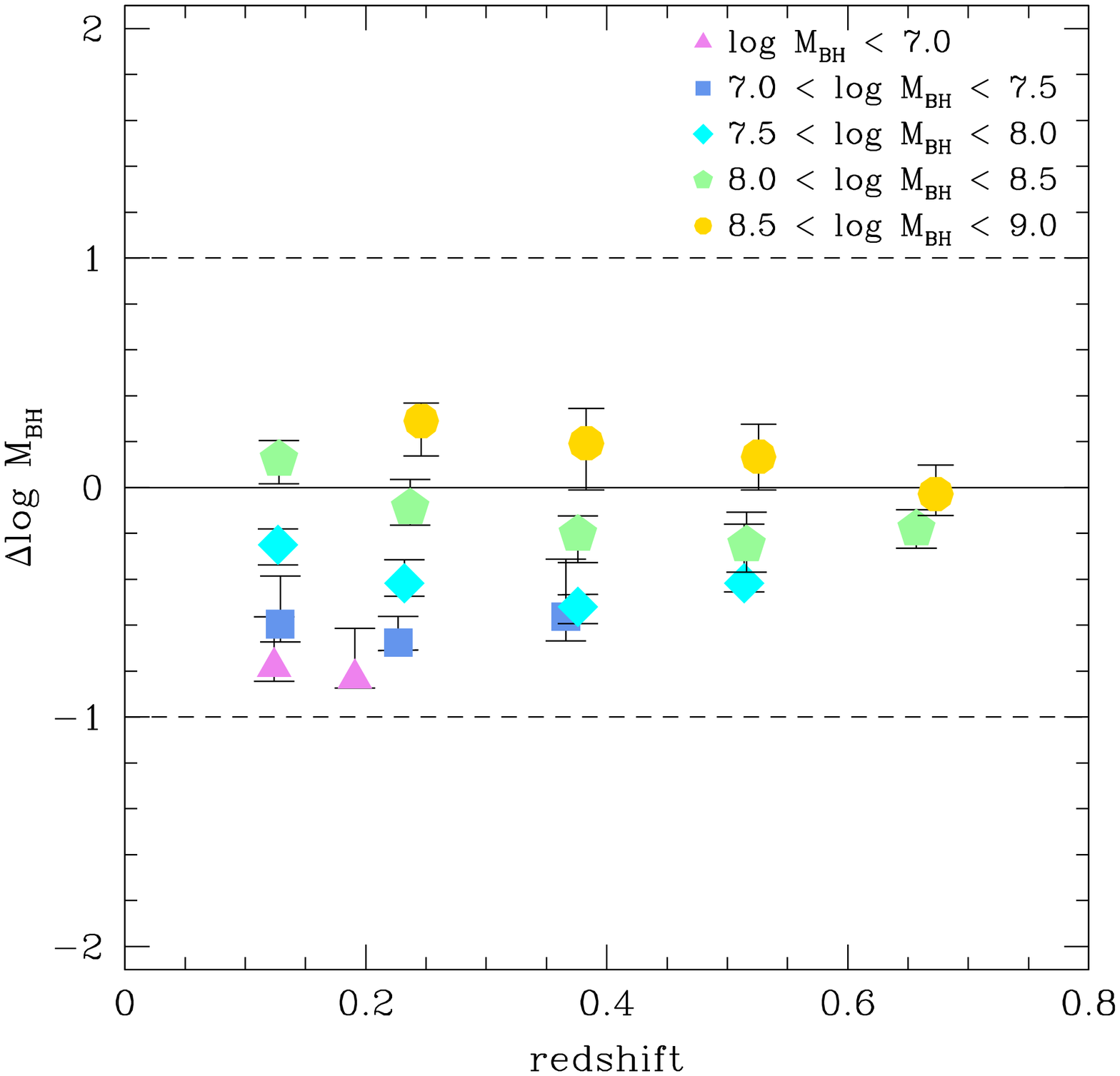}
\caption{Offset of \mbh\ from the local \mbhmgal\ relationship as a function of redshift for the black hole mass bins. }\label{f:dmbh}
\end{figure}

\subsection{The \mbhlgal\ Relation}\label{s:mbhlgal}

Figure \ref{f:mbhmgal} shows the \mbhmgal\ relationship for the mass bins in Table \ref{t:mbins}. (We use a harmonic mean \mbh\ because we are testing agreement with an expected relationship $\mbh \propto \lgal$, and the galaxy luminosity is a harmonic mean.) The solid line is the \mbhmgal\ relationship for locally observed galaxies given by \citet{haering04},
\begin{equation}\label{e:mbhmgal}
\mathrm{log}\, \mbh = 8.20 + 1.12\,\mathrm{log}\, (\mgal/10^{11} \msun),
\end{equation}
in solar units of mass. The trend in Figure \ref{f:mbhmgal} is similar in character to the trend in Figure 3 of S13, which showed the \mbhsigstar\ relationship for the same mass bins. In both cases, the quasar results give a steeper increase in \mbh\ than for the local black hole--bulge relationship. This likely reflects similar selection biases to those causing the departure from the local \mbhsigstar\ relationship in Figure 3 of S13. In addition, it is likely that the lower mass bins contain a substantial number of pseudo-bulges. \citet{kormendy13} show a large fraction of pseudo-bulges for $\mathrm{log}\, \mbh < 7.7$. Finally, the lower mass points in Figure \ref{f:mbhmgal} are likely affected by a disk component, which could displace them by several tenths dex toward higher luminosity compared with the bulge luminosity alone (see Section \ref{s:disk}).

The points in Figure \ref{f:mbhmgal} are mostly displaced to lower \mbh\ or higher \mgal\ relative to the plotted relationship from \citet{haering04}. In contrast, the points in the \mbhsigstar\ plot in Figure 3 of S13 fall mostly above the the reference line from \citet{tremaine02}. Some of this difference may be due to a disk contribution to the host galaxy luminosity. In addition, there is question of possible offsets between the \mbhsigstar\ relationship of \citet{tremaine02} and the \mbhmgal\ relationship in Figure \ref{f:mbhmgal} here. To assess this, we used the adopted Faber--Jackson relationship shown in our Figure \ref{f:fj} to express the \mbhsigstar\ relationship of \citet{tremaine02} in terms of \mbh\ and \lgal. We then used the \citet{magorrian98} expression for the mass-to-light ratio to convert \lgal\ to \mgal, yielding the result $\mbh = 10^{7.92} M_{11}^{0.91}.$ At the relevant \mgal, this is $\sim0.3$~dex lower than the \citet{haering04} relation shown in Figure \ref{f:mbhmgal}. This accounts for part of the difference between Figure 3 of S13 and Figure \ref{f:mbhmgal} here. This issue has little bearing on the redshift-evolution results discussed below and shown in Figure \ref{f:dmbh}, because we compare different redshifts at a given \mbh.

\subsection{Evolution in the \mbhlgal\ Relation with Lookback Time}\label{s:evol}

We assessed the degree to which the quasars in our sample follow the same the \mbhlgal\ relationship as a function of lookback time. In order to do this, we subdivided each of our mass bins into redshift bins, following S13. Bin 1 was $0.1 < z < 0.15$; and the others were incremented by $\Delta z = 0.15$, so that bin 2 was $0.15 < z < 0.30$, etc. We created composite spectra using the objects in each of the redshift bins, and repeated the procedures described in Sections \ref{s:comp} and \ref{s:lgal}. Table \ref{t:mzbins} shows various measured and averaged quantities for the composite spectra for each of these bins. (We omit the 9.25 mass bin because of the uncertainty in \lgal\ noted above.) We took an approach similar to that described in Section 4.2 of S13 to compare the quasar \mbh\ calculated with Equation (\ref{e:mbh}) with the black hole mass inferred from Equation (\ref{e:mbhmgal}) given the mean galaxy luminosity measured for that mass bin, $M_L$. The displacement from the locally determined \mbhlgal\ relationship \citep{haering04} is defined as $\dmbh \equiv \mathrm{log \ } \mbh - \mathrm{\ log \ } M_L$, where $\dmbh = 0$ is perfect agreement with the local \mbhlgal\ relationship and $\dmbh > 0$ indicates \mbh\ is greater than the black hole mass $M_L$ expected for a given \lgal. 

Figure \ref{f:dmbh} shows \dmbh\ versus redshift for all of the mass bins except the highest, for which there were too few objects at each redshift to obtain a measurable \caii\ feature in the composite. Figure \ref{f:dmbh} is similar in character to Figure 4 in S13, which showed the \mbhsigstar\ relationship as a function of time for the same mass bins. The systematic offset from one mass bin to the next likely results from selection effects involving the scatter in the \mbhlgal\ relationship, as discussed in S13. (The discussion in S13 involves the same data and mass bins as here, and their simulations apply equally here. Their simulations were done in terms of \mgal\ and translated into \sigstar\ using $\mgal \propto \sigstar^4$.) In addition, the lower mass points may be displaced downward by a possible disk component to the host galaxies (see below). Overall, there is little apparent evolution in the \mbhlgal\ relationship for the mass bins in Figure \ref{f:dmbh}. There is a general tendency for \dmbh\ to trend downward by about 0.2~dex across the range of redshift represented in Figure \ref{f:dmbh}. A linear least squares fit to our results for the 7.75, 8.25, and 8.75 mass bins gives $d (\dmbh)/dz = -0.36, - 0.44, -0.60$, respectively. Taking account of the dispersion of these values and the errors displayed in Figure \ref{f:dmbh}, we may characterize our results as $d( \dmbh)/dz = -0.4 \pm 0.2$. However, given the various possible biases that can enter at this level, we refrain from assigning significance to the negative trend. Moreover, the \signl\ results in Figure 4 of S13 do not show such a trend. Taken together, our results and those of S13 indicate that the \mbh -- \lgal\ relationship has remained constant within about $\pm0.2~\mathrm{dex}$ since redshift $z \approx 0.7$.

\subsection{Uncertainties}\label{s:uncert}

\subsubsection{AGN Luminosity}

In deriving \mbh\ using Equation (\ref{e:mbh}) above, we have used the full observed luminosity of the AGN, including the host galaxy. Properly, a correction should be made for the host galaxy contribution, resulting in a smaller value for \mbh. This correction is small because the host galaxy makes only a fractional contribution to the total light and because \mbh\ varies only as $L_\mathrm{agn}^{0.5}$. For example, from our fitting of the composites in the $\mathrm{log}\,\mbh = 8.25$ redshift series, the galaxy fraction of the light at $\lambda5100$ decreases from $\sim30\%$ for the lowest redshift bin (8.25.1) to $\sim10\%$ for the highest redshift bin (8.25.5). This corresponds to reduction in \mbh\ (and increase in \dmbh) by 0.08~dex for the lowest redshift and 0.02~dex for the highest redshift. This correction would slightly increase the nominal downtrend in \dmbh\ with increasing redshift in Figure 5, but would not significantly alter our conclusions.

\subsubsection{Template Galaxy}

The template galaxy serves to establish the ratio of the continuum flux at 5100~\AA\ to the flux deficit that constitutes the Ca~K absorption line in the host galaxy spectrum. Use of another template galaxy would yield a different host galaxy luminosity to the extent that this ratio differs from our adopted template. We reanalyzed the 7.75.3 composite spectrum using a different galaxy from the \citet{bernardi08} sample, SDSS J094035.88+022949.9 (spSpec-52026-0477-504). This is more luminous than the adopted template, but still within the range of host galaxy luminosities in our results. The host galaxy luminosity derived using this template was lower by 0.07~dex than the value for our adopted template. As a further test, we examined the depth of Ca K relative to the 5500~\AA\ flux in the SDSS ``early type galaxy'' cross-correlation template (spDR2-023)\footnote{http://http://classic.sdss.org/dr7/algorithms/spectemplates/}, which is a composite of many SDSS galaxies. This template agrees with the adopted template within a few percent in the salient flux ratio.

\subsubsection{Sample Size}\label{s:sample}

The number of objects in the extreme mass bins is much smaller than for the intermediate mass bins. As a test of the sensitivity of our procedure to the object count, we formed random subsets of 100 objects from the 7.75 mass bin and carried out the template fitting procedure. The results for \lgal\ showed a spread of $\pm0.05$~dex around the result for the full number of objects in this bin. This gives an indication of the uncertainty from sample statistics for our smaller bins
(see Tables \ref{t:mbins} and \ref{t:mzbins}).

\subsubsection{Stellar Population}\label{s:population}

One potential systematic error involves the stellar population of the host galaxies. This is a function of redshift and possibly of \mbh. Our redshift bins have central values ranging from $z = 0.125$ to $z = 0.675$, with look-back times of 1.6 to 6.2~Gyr. We have used a single, low redshift galaxy template to subtract the \caii\ absorption features from the quasar composite spectra and thereby measure the galaxy component in the spectra. If the actual combined spectrum of the quasar host galaxies has, for example, a smaller equivalent width (EW) for the calcium feature, we will underestimate the galaxy contribution to the composite spectrum. If the higher redshift galaxies have a different EW of \caii, this could give a spurious evolutionary trend in the ratio of \mbh\ to \lgal. For a rough estimate of this effect, we examined the stellar population synthesis models\footnote{http://people.ucsc.edu/{\textasciitilde}conroy/CvD12.html} of \citet{conroy12}, choosing for simplicity the single-age models with solar abundances and a Salpeter initial mass function. For ages 3, 6, and 11~Gyr, respectively, we measured an EW for the combined \caii\ feature of 19.4, 19.5, and 23.6~\AA. Consider a model in which all star formation occurred in a single burst at $z=2$, a lookback time of 10.2~Gyr. Then the synthetic spectra suggest that we may have underestimated the continuum at $\lambda3950$ in the quasar spectra by 0.022 and 0.083~dex at $z = 0.125$ and $z = 0.675$, respectively. (We have scaled the logarithm of the EW ratios of the population models linearly in elapsed time based on the three ages quoted above.) The differential effect is to suggest that we have underestimated the $F_\lambda(3950)$ by $\sim0.061$~dex at $z = 0.675$ relative to $z = 0.125$.

An offsetting effect comes from the evolving color of the galaxy continuum. We used a fixed ratio of $F_\lambda(5500)/F_\lambda(3950)$ based on the galaxies observed by \citet{salviander08}; the adopted ratio agrees well with the SDSS template galaxy used here. However, if the higher redshift galaxies have younger and bluer stellar populations, then the ratio of $F_\lambda(5500)/F_\lambda(3950)$ is less than we assumed. Thus we should lower the $\lambda5500$ continuum for the higher redshift composites, offsetting the EW effect. From the Conroy \& van Dokkum spectra, we measure $F_\lambda(5500)/F_\lambda(3950) = 1.6, 1.9, \mathrm{and}\ 2.2$ for ages 3, 6, and 11~Gyr, respectively. A log-linear interpolation suggests that our procedure overestimates $F_\lambda(5500)$ by 0.016~dex at $z = 0.125$ and by 0.093 dex at $z = 0.675$. Thus, the color effect by itself causes us to overestimate $F_\lambda(5500)$ by $\sim 0.077$~dex at $z = 0.675$ relative to $z = 0.125$.

The combined effect of the EW and color evolution is to underestimate $F_\lambda(5500)$ by 0.006~dex at $z = 0.125$ and to overestimate it by 0.016 at $z = 0.675$. The differential effect is to overestimate $F_\lambda(5500)$ at $z = 0.675$ by 0.022~dex relative to $z=0.125$, which is an insignificant amount that is within the uncertainties of our analysis. We conclude that these stellar evolution effects do not seriously bias our results.
 
 \subsubsection{Fiber Size}\label{s:fiber}
 
 Another concern is the loss of host galaxy light falling outside the 3~arcsec diameter of the light fibers used in the SDSS spectrograph. For our redshift bins, the central redshift is
 (0.125, 0.225, 0.375, 0.525, 0.675). The 1.5~arcsec fiber radius corresponds to (3.4, 5.4, 7.7, 9.4, 10.6~kpc), respectively. Adjusted to $\hnot = 70$, the results of \citet{bender92} give an average effective radius $r_e \approx 3~\mathrm{kpc}$ for ellipticals and bulges within $\pm0.5$ mag of $M_\mathrm{V} = -21.15$. (The Virgo cluster sample of \citet{kormendy09} suggests $r_e$ larger by 0.1 or 0.2~dex.) For a simple scaling $r_e \propto L_\mathrm{V}$, the mean luminosity in the respective redshift bins predicts an effective radius of (1.4, 2.3, 3.6, 3.8, 4.1~kpc) or (0.62, 0.62, 0.74, 0.62, 0.58~arcsec) for the mass bin centered on $\mathrm{log}\, \mbh = 8.25$. The SDSS spectroscopic survey has a median effective seeing of 1.43~arcsec \citep{stoughton02}. For a rough estimate, we take the radius containing half the light in the point-spread function (PSF) to be $S_{\mathrm e} = 0.5\times \mathrm{FWHM} = 0.72~\mathrm{arc sec}$, based on a Gaussian PSF. Adding this in quadrature to the above angular radii, we find an image effective radius of (0.96, 0.98, 1.02, 0.96, 0.94~arcsec) for the five redshift bins. This corresponds to a fiber light-capture fraction $\lfib/\ltot$ of (0.62, 0.62, 0.61, 0.62, 0.63) for a \citet{devauc48} $r^{1/4}$ light profile or (0.81, 0.80, 0.78, 0.82, 0.83) for a Gaussian profile. These estimates indicate that light lost outside the SDSS fibers is significant, but that differences in the capture fracture across the redshift bins is not a serious uncertainty.

 \citet{falomo14} studied the host galaxies of quasars in SDSS Stripe 82. Their sample is comprised of 416 objects in the redshift range $0.1 < z < 0.5$, with a mean of $z = 0.39$. From model fitting, Falomo et al. succeeded in obtaining the host luminosity \lgal\ and effective radius (half-light radius) \reff\ for a majority of the objects. We created a composite spectrum for the 305 objects with measured host properties, and determined from template subtraction a harmonic mean host galaxy luminosity in $V$ of $\mathrm{log}\,\lgal = 10.60$ in solar units (see procedure above). For comparison, the harmonic mean of the luminosity found by Falomo et al. for the same objects is $\mathrm{log}\,\lgal = 10.67$ in $V$, where we have used $V - R = 0.7$ following Falomo et al. This agreement is good in the context of the above estimate of the fiber light capture. We also carried out our composite procedure for the $z = 0.36$ objects of \citet{bennert10}. Our procedure yielded a luminosity of $10^{10.56}$ \lsun\ before evolutionary correction. The harmonic mean of the total host luminosity is $10^{10.57}~\lsun$ for the Bennert et al. objects after a correction of 0.17 dex for fiber light capture based on the individual effective radii quoted by Bennert et al. and including seeing. These comparisons support the validity of our procedure at the level of $\sim0.1$~dex accuracy.

\subsubsection{Disk Contribution}\label{s:disk}
 
\citet{kormendy13} summarize evidence that black hole masses show little correlation with galactic disks. \citet{sanghvi14} found a substantial disk component for most of the host galaxies in an imaging study of a sample of SDSS quasars with \mbh\ in the range $10^7$ to $10^{8.3}~\msun$ and $0.5 < z < 1.0$. Combining their results with other studies over a large range of \mbh, they found that the objects with $\mbh < 10^{8.2}$ fell below the \mbh\ -- \mgal\ relationship obeyed by more massive black holes. Sanghvi et al. attributed this to the presence of a substantial disk component for the hosts of the smaller black holes, correction for which brought them into agreement with the trend for the larger black holes. Their analysis suggests a disk/total luminosity ratio of as much as $\sim0.5$~dex for the hosts of $10^{7.75}~\msun$ black holes. Such a correction could account for much of the departure of our lower mass bins from the local \mbh\ -- \mbulge\ in Figure \ref{f:mbhmgal} and contribute to the vertical depression of the lower bins in Figure \ref{f:dmbh}. However, for the redshift trends presented in Figure \ref{f:dmbh}, we have binned the objects by black hole mass. We assume that the trends with redshift within a given bin in \mbh\ are not seriously affected by different disk contributions to the different redshift bins at a given black hole mass.
 
\section{DISCUSSION}\label{s:discuss}

The conclusion of this work is that there is little evolution in the \mbhlgal\ relationship for SDSS quasars over the redshift range $z = 0.1$ to 0.7. This resembles the findings of S13 for the \mbhsigstar\ relationship as inferred from \oiii\ width as a surrogate for \sigstar. At a level of $\pm0.2$ in \dmbh, host galaxies and their black holes have maintained the present day proportionality at least since redshift 0.7.

\citet{woo2006,woo2008} report an offset $\dmbh = +0.50 \pm 0.22 \pm 0.25$ at $z = 0.57$ and a similar offset at $z = 0.36$, based on measurements of \mbh\ and \sigstar\ in a sample of active galaxies. Similar results based on host galaxy luminosities derived from {\it Hubble Space Telescope} imaging are reported by \citet{treu07} and \citet{bennert10}. Bennert et al. find a dependence $\mbh/\lbulge \propto (1+z)^{2.8}$, or $\mbh/\lbulge \propto (1+z)^{1.4}$, when including higher redshift measurements from the literature. Over our redshift range, a dependence $\mbh/\lbulge \propto (1+z)^{2.8}$ corresponds to $d(\dmbh)/dz \approx +0.8$. Such a slope is inconsistent with our results in Figure \ref{f:dmbh}. However, our results refer to the total host galaxy luminosity, including any disk component whose light falls within the SDSS fiber diameter. Bennert et al. find little evolutionary trend when comparing \mbh\ to total host luminosity for their objects at redshift 0.36 and 0.57. They suggest that at redshifts below $z \approx 1$, the host evolution may largely involve redistribution of existing stars into the bulge component.

\citet{peng06} examined the black hole--bulge luminosity relationship for a sample of lensed and unlensed quasars spanning a wide range of redshift. For $z > 1.7$, they find that the ratio $\mbh/M_*$ of black hole mass to host stellar mass is larger by a factor $\sim4$, relative to the present. For $1 < z < 1.7$, they find that $\mbh/M_*$ is at most a factor of two larger than today, and is consistent with no evolution. \citet{decarli10} present results for a sample of quasars with redshift up to $z = 3$. Considering the entire range of redshift, they find \dmbh\ increasing by $\sim 0.3$ per unit redshift. However, for the subset of their objects having a nucleus/host luminosity ratio less than 5, their low redshift data by themselves give little evidence for significant evolution in \dmbh\ between redshifts 0.4 and 0.7. These results appear to be consistent with our conclusion of little evolution since redshift $z = 0.7$.

An absence of significant evolution in the \mbhlgal\ and \mbhsigstar\ relationships out to $z = 0.7$ is found in this work and S13. Combined with evidence of a larger ratio of \mbh\ to spheroid luminosity for luminous AGNs at redshifts of two and greater, this is consistent with a scenario in which black holes grew rapidly in the early universe, and host galaxy spheroids catch up by $z \approx 1$ \citep{kormendy13}.

\acknowledgments

We thank the anonymous referee for valuable suggestions that substantially improved this paper. We thank K. Gebhardt for valuable advice and for the use of his spectrum fitting and compositing programs. We thank C. Conroy for helpful discussions. An early version of this study was begun while EWB was a postdoctoral fellow at LUTH, Observatoire de Paris, supported by Marie Curie Incoming European Fellowship contract MIF1-CT-2005-008762 within the 6th European Community Framework Programme. Funding for the Sloan Digital Sky Survey (SDSS) has been provided by the Alfred P. Sloan Foundation, the Participating Institutions, the National Aeronautics and Space Administration, the National Science Foundation, the U.S. Department of Energy, the Japanese Monbukagakusho, and the Max Planck Society. The SDSS Web site is http://www.sdss.org/. The SDSS is managed by the Astrophysical Research Consortium (ARC) for the Participating Institutions. The Participating Institutions are The University of Chicago, Fermilab, the Institute for Advanced Study, the Japan Participation Group, The Johns Hopkins University, the Korean Scientist Group, Los Alamos National Laboratory, the Max-Planck-Institute for Astronomy (MPIA), the Max-Planck-Institute for Astrophysics (MPA), New Mexico State University, University of Pittsburgh, University of Portsmouth, Princeton University, the United States Naval Observatory, and the University of Washington.

\section{Appendix}\label{s:appendix}

The composite spectra were composed as described in Section \ref{s:comp}. This involves normalizing the individual quasar spectra by dividing the specific flux \flam\ by the wavelength-averaged flux $x_c = \, \langle F_\lambda \rangle $ for a given quasar. The recovery of the average luminosity of the host galaxies contributing to the composite involves reversing this normalization to get the actual host galaxy flux in the composite, and using the luminosity distance $d_L$ to get the luminosity from the flux. Consider the special case in which all the host galaxies have the same luminosity but lie at different distances from earth. The luminosity distance is defined such that the specific luminosity at rest wavelength $\lambda_1$ is given by
$\lambda_1 L_{\lambda_1} = 4\pi d_L^2\, \lambda_2 F_{\lambda_2}$,
where $\lambda_2 = (1+z) \lambda_1$. The galaxy flux in the composite spectrum is given by
\begin{equation}\label{e:fcomp}
 f_{\lambda,\mathrm{gal}}^c = \langle F_{\lambda,\mathrm{gal}}/x_c \rangle = \langle (L_{\lambda,\mathrm{gal}}/4\pi d_L^2) [(x_c \, (1+z)]^{-1} \rangle.
\end{equation}
Here $\lambda$ refers to the rest wavelength of interest, taken to be 3950~\AA\ in our work, and $F_{\lambda}$ is the received specific flux at $(1+z)\lambda$. Factoring out the constant $L_{\lambda,\mathrm{gal}}$ and solving, we find Equation (\ref{e:lbar}).

Now consider the meaning of $\bar{L}_{\lambda,\mathrm{gal}}$ when Equation (\ref{e:lbar}) is applied to the general case with a range of luminosity and redshift for the quasars contributing to a given composite. Using the index $i$ to label the individual quasars, we have
$4\pi\,d_{L,i}^2\,(1+z_i) = L_{\lambda,\mathrm{gal},i}/F_{\lambda,\mathrm{gal},i}$.
With this, we rewrite Equation (\ref{e:lbar}) as
\begin{equation}\label{e:lbarx}
\bar{L}_{\lambda,\mathrm{gal}} = \, f_{\lambda,\mathrm{gal}}^c \,\langle [(L_{\lambda,\mathrm{gal}}/F_{\lambda,\mathrm{gal}}) x_c]^{-1}\rangle^{-1}.
\end{equation}
Noting that $ f_{\lambda,gal}^c$ may be taken inside the summation giving the average, we define
\begin{equation}\label{e:eta}
\eta_i \equiv f_{\lambda,\mathrm{gal}}^c /(F_{\lambda,\mathrm{gal},i}\, x_{c,i}^{-1}) = \, \langle F_{\lambda,\mathrm{gal},i}\, x_{c,i}^{-1}\rangle/(F_{\lambda,\mathrm{gal},i}\, x_{c,i}^{-1}).
\end{equation}
Using this in Equation (\ref{e:lbarx}), we have
\begin{equation}\label{e:lharm}
\bar{L}_{\lambda,\mathrm{gal}} = \, \langle (L_{\lambda,\mathrm{gal}} \, \eta)^{-1}\rangle^{-1}.
\end{equation}

Recalling that $x_c$ is the wavelength-averaged specific flux of a given quasar (galaxy plus AGN), we see that $F_{\lambda,\mathrm{gal}}/x_c$ is approximately the galaxy fraction of the total specific flux of the quasar, at the fiducial wavelength. If this were uncorrelated with $L_{\lambda,\mathrm{gal}}$, then $\bar{L}_{\lambda,\mathrm{gal}}$ would be the harmonic mean of the individual galaxy luminosities. In practice, however, the galaxy fraction decreases with increasing quasar luminosity, roughly a factor of three over a range of two orders of magnitude in luminosity. Our composites each have quasars with a wide range of luminosity, but typically the central half of the objects span a range of 0.5~dex in luminosity. Consider a toy model with a composite composed of only two quasars with luminosity $L_2 = 10^{0.5}L_1$ and galaxy fraction $\eta_2 = 10^{-0.125} \eta_1$. The harmonic mean luminosity is $1.5 L_1$, but Equation (\ref{e:lharm}) gives $\bar{L}_{\lambda,\mathrm{gal}} = 1.40 L_1$. Thus, the algorithm underestimates the harmonic mean by 7\%. However, this bias should be similar between the various composites, which have a similar distribution of individual quasar luminosities. Since our main goal is to assess the evolution of the black hole--galaxy relation by comparing results for composites with different black hole mass, we will omit any correction for this bias and assume that Equation (\ref{e:lbar}) gives the harmonic mean galaxy luminosity for a given composite.


\begin{thebibliography}{}

\bibitem[Abazajian et al.(2009)]{abazajian09} Abazajian, K.~N., 
Adelman-McCarthy, J.~K., Ag{\"u}eros, M.~A., et al.\ 2009, \apjs, 182, 543

\bibitem[Bender et al.(1992)]{bender92} Bender, R., Burstein, D., \& Faber, S. M.\ 1992, \apj, 399, 462

\bibitem[Bennert et al.(2010)]{bennert10} Bennert, V. N., et al.\ 2010, \apj, 708, 1507

\bibitem[Bernardi et al.(2008)]{bernardi08} Bernardi, M., Fritz, A., Hyde, J. B., Sheth, R. K., Gebhardt, K., \& Nichol, R. C.
2008, MNRAS, 391, 1191

\bibitem[Bonning et al.(2005)]{bonning05} Bonning, E.~W., 
Shields, G.~A., Salviander, S., \& McLure, R.~J.\ 2005, \apj, 626, 89

\bibitem[Conroy \& van Dokkum(2012)]{conroy12} Conroy, C., \& van Dokkum, P.\ 2012, \apj, 474, 69

\bibitem[Decarli et al.(2010)]{decarli10} Decarli, R., Falomo, R., Treves, A., Labita, M., Kotilainen, J. K., \& Scarpa, R. 2010, \mnras, 402, 2453

\bibitem[de Vaucouleurs(1948)]{devauc48} de Vaucouleurs, G. 1948, Ann. Astrophys., 11, 247

\bibitem[Falomo et al.(2014)]{falomo14} Falomo, R., Bettoni, D., Karhunen, K., Kotilainen, J. K., \& Uslenghi, M. 2014, \mnras, 440, 476


\bibitem[Ferrarese \& Merritt(2000)]{ferrarese00} Ferrarese, L., \& Merritt, D.\ 2000, \apjl, 539, L9

\bibitem[Gebhardt et al.(2000)]{gebhardt00} Gebhardt, K., Bender, R., Bower, G., et al.\ 2000, \apjl, 539, L13

\bibitem[Greene \& Ho(2006a)]{greene06oiii} Greene, J. E., \& Ho, L. 2006a, \apj, 627, 721

\bibitem[Greene \& Ho(2006b)]{greene06sig} Greene, J. E., \& Ho, L. 2006b, \apj, 641, 117

\bibitem[H\"aring \& Rix(2004)]{haering04} H\"aring, N., \& 
Rix, H.-W. 2004, \apjl, 604, L89

\bibitem[Kormendy \& Bender(2013)]{kormendy13fj} Kormendy, J., \& Bender, R. 2013, \apjl, 769, L5

\bibitem[Kormendy et al.(2009)]{kormendy09} Kormendy, J., Fisher, D. B., Cornell, M. E., \& Bender, R.\ 2009, \apjs, 182, 216

\bibitem[Kormendy \& Ho(2013)]{kormendy13} Kormendy, J., \& Ho., L. C. 2013, \araa, 51, 511

\bibitem[Magorrian et al.(1998)]{magorrian98} Magorrian, J., et 
al.\ 1998, \aj, 115, 2285 

\bibitem[Onken et al.(2004)]{onken04} Onken, C.~A., et al. 2004, \apj, 615, 645

\bibitem[Osterbrock \& Ferland (2006)]{osterbr06} Osterbrock, D.~E., \& Ferland 2006, 
`Astrophysics of Gaseous Nebulae and Active Galactic Nuclei,' 2nd ed., University Science Books 

\bibitem[Peng et al.(2006)]{peng06} Peng, C.~Y., Impey, C.~D., 
Rix, H.-W., et al.\ 2006, \apj, 649, 616

\bibitem[Salviander et al.(2007)]{salviander07} Salviander, S., 
Shields, G.~A., Gebhardt, K., \& Bonning, E.~W.\ 2007, \apj, 662, 131 (S07)

\bibitem[Salviander et al.(2008)]{salviander08} Salviander, S., 
et al.\ 2008, \apj, 687, 828

\bibitem[Salviander et al.(2013)]{salviander13} Salviander, S., \& Shields, G.~A.\ 2013, \apj, 764, 80 (S13)

\bibitem[Sanghvi et al.(2014)]{sanghvi14} Sanghvi, J., Kotilainen, J. K., Falomo, R., Decarli, R., Karhunen, K., \& Uslenghi, M. 2014, \mnras, in press [arXiv:1409.1948]

\bibitem[Shields et al.(2003)]{shields03} Shields, G.~A., 
Gebhardt, K., Salviander, S., Wills, B.~J., Xie, B., Brotherton, M.~S., 
Yuan, J., \& Dietrich, M.\ 2003, \apj, 583, 124 (SO3)

\bibitem[Shields et al.(2006a)]{shields06leiden} Shields, G.~A., 
Salviander, S., \& Bonning, E. W. 2006a, New Astronomy Reviews, 50, 809

\bibitem[Shields et al.(2006b)]{shields06co} Shields, G.~A., 
Menezes, K.~L., Massart, C.~A., \& Vanden Bout, P.\ 2006b, \apj, 641, 683

\bibitem[Stoughton et al.(2002)]{stoughton02} Stoughton, C. 2002, \aj, 123, 485

\bibitem[Tremaine et al.(2002)]{tremaine02}
{Tremaine}, S., {Gebhardt}, K., {Bender}, R., {Bower}, G., {Dressler},
A., {Faber}, S.~M., {et al.} 2002, \apj, 574, 740

\bibitem[Vanden Berk et al.(2001)]{vandenberk01} Vanden Berk, D.~E., 
et al.\ 2001, \aj, 122, 549

\bibitem[van Dokkum \& Franx(2001)]{vandokkum01} van Dokkum, P. G., \& Franx, M.\ 2001, \apj, 553, 90

\bibitem[Treu et al.(2007)]{treu07} Treu, T., Woo, J.-H., 
Malkan, M.~A., \& Blandford, R.~D.\ 2008, \apj, 667, 117

\bibitem[Woo et al.(2008)]{woo2008} Woo, J.-H., Treu, T., 
Malkan, M.~A., \& Blandford, R.~D.\ 2008, \apj, 681, 925

\bibitem[Woo et al.(2006)]{woo2006} Woo, J.-H., Treu, T., 
Malkan, M.~A., \& Blandford, R.~D.\ 2008, \apj, 645, 900

\end{thebibliography}
\end{document}